\documentclass[journal,11pt,onecolumn]{IEEEtran}
\usepackage[top=1in, bottom=1in, left=1.25in, right=1.25in]{geometry}
\usepackage[doublespacing]{setspace}
\usepackage{amsmath}
\usepackage{booktabs}
\usepackage{array}
\usepackage[table]{xcolor}
\DeclareMathOperator*{\argmax}{argmax}
\newcommand{\ma}[1]{{\bf #1}}

\ifCLASSINFOpdf
\else
\fi
\usepackage{amsmath}
\usepackage{amssymb}
\usepackage{xcolor}
\usepackage{url}
\usepackage{bbm}
\usepackage{array}
\usepackage{verbatim}
\usepackage{mdframed}
\usepackage{caption}

 \usepackage{graphicx}

\usepackage{todonotes}

\begin{document}
%
\title{Community-Aware Graph Signal Processing}
%
%
%

\author{Miljan~Petrovi\' c$^\star$,~Rapha\"el~Li\'egeois$^\star$,~Thomas~A.~W.~Bolton,~\textit{Student~Member},~and Dimitri~Van~De~Ville,~\textit{Fellow}
\thanks{$^\star$contributed equally. \newline M. Petrovi\' c, R. Li\'egeois, T. A. W. Bolton and D. Van De Ville are with the Institute of Bioengineering, \'{E}cole Polytechnique F\'ed\'erale de Lausanne, Campus Biotech, Geneva, Switzerland, and the Department of Radiology and Medical Informatics, University of Geneva, Geneva, Switzerland. T. A. W. Bolton is also with the Department of Decoded Neurofeedback, ATR Computational Neuroscience Laboratories, Soraku-gun, Kyoto, Japan.}}
%
%
\markboth{Community-Aware Graph Signal Processing}%
{Shell \MakeLowercase{\textit{et al.}}: Community-Aware Graph Signal Processing}
%
\maketitle
\vspace{-2.5cm}
\begin{abstract}
The emerging field of graph signal processing (GSP) allows to transpose classical signal processing operations (e.g., filtering) to signals on graphs. The GSP framework is generally built upon the graph Laplacian, which plays a crucial role to study graph properties and measure graph signal smoothness. Here instead, we propose the graph modularity matrix as the centerpiece of GSP, in order to incorporate knowledge about graph community structure when processing signals on the graph, but without the need for community detection. We study this approach in several generic settings such as filtering, optimal sampling and reconstruction, surrogate data generation, and denoising. Feasibility is illustrated by a small-scale example and a transportation network dataset, as well as one application in human neuroimaging where community-aware GSP reveals relationships between behavior and brain features that are not shown by Laplacian-based GSP. This work demonstrates how concepts from network science can lead to new meaningful operations on graph signals.
\end{abstract}


%
\IEEEpeerreviewmaketitle

\section{Introduction}
%
%
%
%

Network science is a multidisciplinary field that accounts for complex structure of data, providing new interpretations of datasets in diverse scientific disciplines ranging from humanities to physics and biomedicine. Naturally, analysis of network data relies on methods from graph theory, but also from statistical mechanics, statistical inference, advanced visualization, and domain knowledge from applied fields. More recently, graph signal processing (GSP) emerged as a new research theme at the intersection between signal processing and graph theory, with a particular focus on processing \emph{graph signals} that associate values to the nodes of the graph. In many cases, the graph Fourier transform was defined by the eigendecomposition of the graph Laplacian; i.e., the eigenvectors of the Laplacian are considered as graph Fourier basis vectors, and the associated eigenvalues are graph frequencies~\cite{Shuman2013}. Such graph Fourier transform can then generalize various classical signal processing tools to graphs~\cite{Shuman2013,Sandryhaila2013}, such as the wavelet transform~\cite{Hammond2011}, as well as theoretical considerations about graph uncertainty principles~\cite{Tsitsvero.2016}. 

The graph Laplacian defines the second-order derivative on the graph and is therefore linked to smoothness, but alternative operators can explore other properties of graphs and graph signals. For example, \emph{community structure} is a particularly interesting concept from network science where nodes inside a community are more strongly interconnected than with the rest of the graph~\cite{Fortunato2016,Newman2004}. Community structure turned out to be present and relevant for a broad range of applications in sociology~\cite{Jeub2015}, transportation~\cite{Guimera2005}, biology~\cite{Guimera2005nature} or neuroscience~\cite{Meunier2010}. In practice, communities can be found by maximizing the modularity index that evaluates the density of connections within clusters against a \emph{degree-matched graph} where no cluster preference exists~\cite{Newman2006}. Similar to Laplacian-based spectral clustering, where the Laplacian eigenvectors with smallest non-zero eigenvalues are considered since they optimize the convex relaxation of the graph cut criterion~\cite{vonLuxburg2007}, one approach for community detection is to compute the eigendecomposition of the modularity operator and consider the eigenvectors with largest eigenvalues~\cite{Fortunato2016}.

In this paper, we set the foundations for community-aware GSP by introducing the modularity operator at the heart of the framework. This allows to define GSP operations that are aware of the graph community structure, but without the need of explicit community detection. After recalling basic GSP notions (Section \ref{sec:GSP}), we define the modularity index and corresponding operator, highlighting the differences with the Laplacian (Section \ref{sec:Mod}). We then detail how GSP operations such as filtering, sampling, and denoising, can be rendered community aware (Section \ref{sec:GSP_M}). Using the OpenFlights and a functional magnetic resonance imaging (fMRI) datasets, we illustrate the benefits of community-aware GSP over its Laplacian-based counterpart (Sections \ref{sec:GSP_M} and \ref{sec:Val}).

\section{Graph Signal Processing}
\label{sec:GSP}
We consider an undirected graph $\mathcal{G}=(\mathcal{N},\mathcal{E})$ with node set $\mathcal{N}$ of cardinality $N$ and edge set $\mathcal{E}$. $\mathcal{G}$ can also be represented by the $N\times N$ weighted adjacency matrix $\ma A$, whose entry $a_{i,j}$ is non-zero and indicates the edge weight for an edge $(i,j)\in\mathcal{E}$ that runs from node $i$ to node $j$. For an undirected graph, $\ma A$ is symmetric; i.e., it holds that $a_{i,j}=a_{j,i}$ and $\ma A=\ma A^\top$. We will refer to a subgraph $\mathcal{G}_S=(\mathcal{N}_S,\mathcal{E}_S)$ by its node set $\mathcal{N}_S\subset\mathcal{N}$ and assume $\mathcal{E}_S$ containing all edges $(i,j)$ between nodes in $\mathcal{N}_S$. A graph signal associated to $\mathcal{G}$ is a vector $\ma x\in \mathbb{R}^N$ that attributes values $x_i$ to the nodes $i=1,2,\ldots,N$. The neighborhood of a node $i$ is defined as the set of nodes $\mathcal{N}_i$ connected to it. A graph shift operator is defined as a linear operator on the space of signals, such that each entry of the shifted graph signal is a linear combination of input signal values, which often only involves neighboring entries to the one at hand~\cite{Sandryhaila2013}. Therefore, the shift operator can be represented by a symmetric matrix $\ma S\in \mathbb{R}^{N \times N}$ that associates weights $s_{i,j}$ to edges $(i,j)$ such that $\ma x_\text{shift}=\ma S \ma x$. We will consider graph operators $\mathcal{H}$ that are shift-invariant under $\ma S$ and thus satisfy $\mathcal{H}\ma S\ma x=\ma S \mathcal{H}\ma x$ and can be represented as a matrix polynomial of $\ma S$~\cite{Ortega2018}; i.e., $\mathcal{H}=p(\ma S)=\sum_{k=0}^K h_k \ma S^k$, with maximum degree of $N-1$ due to the Cayley-Hamilton theorem.

The eigendecomposition of the shift operator provides the factorization
\begin{equation}
\label{Eq:Eig}
\ma S = \ma U \ma \Lambda \ma U^{\top},
\end{equation}
where $\ma U = [\ma u_1, \ldots, \ma u_{N}]$ contains the $N$ eigenvectors and $\ma \Lambda = \text{diag} (\lambda_1, \ldots, \lambda_{N})$ is a diagonal matrix with the corresponding eigenvalues. This allows to write the graph operator $\mathcal{H}$ alternatively as $\mathcal{H}=p(\ma S)=\ma U \text{diag}(\tilde{\ma h})\ma U^{\top}=\ma U\tilde{\ma H}\ma U^\top$, where the entries $\tilde{h}_i=p(\lambda_i)=\sum_{k=0}^K h_k\lambda_i^k$ of $\tilde{\ma H}$ yield the spectral characterization of the graph operator. For the perspective of GSP, a given shift operator $\ma S$ defines the Graph Fourier transform (GFT) of the graph signal $\ma x$ as~\cite{Shuman2013}:
\begin{equation}
\hat{\ma x} =  \ma U^{\top} \ma x, \  \text{ and } \ \ma x =  \ma U  \hat{\ma x},
\end{equation}
where $\ma U$ is defined as in Eq.\,\eqref{Eq:Eig} and $\hat{\ma x}$ contains the spectral coefficients of the GFT. The graph operator $\mathcal{H}$ can then be implemented elegantly in the graph Fourier domain as 
\begin{equation}
\label{eq:filtering}
    {\ma x}_\text{out}= \mathcal{H} \ma x= p(\ma S) \ma x = \ma U p(\ma\Lambda) \ma U^{\top} \ma x = \ma U \tilde{\ma H}\hat{\ma x},
\end{equation}
which allows to directly specify $\tilde{\ma H}$ in terms of a spectral window (e.g., low-pass, band-pass, high-pass) for graph filtering operations~\cite{Ortega2018}. Beyond filtering, other operations have been extended to the graph domain, such as stationarity analysis~\cite{Marques2017}, wavelet transforms~\cite{Hammond2011}, or convolutional neural networks~\cite{Defferrard2016}.

One common choice for $\ma S$ is the weighted graph adjacency matrix $\ma A$~\cite{Sandryhaila2013, Huang2018}. Another one is the Laplacian matrix $\ma L=\ma D-\ma A$, where $\ma D=\text{diag}(k_1,k_2,\ldots,k_N)$ is the degree matrix with $k_i=\sum_{j=1}^{N} a_{i,j}$ the weighted degree~\cite{Hammond2011, leonardi1302}. For the latter, the eigenvalues are sometimes referred to as \emph{graph frequencies} and reflect smoothness in terms of the signal variation norm of the corresponding eigenvectors~\cite{Ortega2018}. For a graph signal $\ma x$, its smoothness is measured by the quadratic form
\begin{equation}
\label{eq_quadvarL}
q_L(\ma x)= \sum_{i\neq j} a_{i,j}(x_i-x_j)^2 = \ma x^\top \ma L\ma x = \sum_{i=1}^N \lambda_i \hat{x}_i^2,
\end{equation}
which shows that measuring smoothness in the spectral domain can be done by weighting with the graph frequencies. 
The example in Fig.~\ref{fig_eigvecs_toy} illustrates the Laplacian eigendecomposition for a simple graph and will be discussed in more details later.

\section{Community structure}
\label{sec:Mod}
Communities refer to dense subgraphs $\mathcal{P}_c\subset\mathcal{N}$, $c=1,\ldots,C$, that are well separated from each other, and manifested at the ``mesoscale'' level between local nodal and global graph properties~\cite{Fortunato2016}. A large number of measures have been proposed with the purpose to discover community structure of an observed graph. For our aim, it is insightful to first revisit the graph Laplacian as it relates to one aspect of community structure, which is quantifying the separation between subgraphs. Specifically, the splitting of a graph into two mutually exclusive subgraphs $\mathcal{P}_1$ and $\mathcal{P}_2$ can be encoded by a vector $\ma s$ whose entries $s_i=+1$ or $-1$ indicate whether a node $i$ belongs to the first or second subgraph, respectively. The graph cut size---number of connections running between the two subgraphs---can then be related to the Laplacian as 
\begin{equation}
\label{Eq:CutSize}
R = \frac{1}{2}\sum_{\substack{i,j\\s_i\ne s_j}} a_{i,j} = \frac{1}{2}\sum_{i,j} \left(\frac{1-s_i s_j}{2}\right) a_{i,j} = \frac{1}{4}\ma s^\top \ma L\ma s.
\end{equation}
Optimizing $R$ by convex relaxation of $\ma s$ (i.e., allowing the entries to take any value) leads to the well-known spectral clustering~\cite{vonLuxburg2007}. The eigenvector of $\ma L$ with smallest non-zero eigenvalue (a.k.a.~Fiedler vector) provides the solution to the bipartition problem. Recent work has also used graph wavelets to enable multiscale subgraph discovery~\cite{Tremblay.2014}.

The network-science view on community structure considers the adjacency matrix as a realization of an underlying stochastic model that defines edge probabilities within and between subgraphs. Stochastic block models (SBMs)~\cite{Lee2019} are the best known generative models that can express assortativity (preferential connectivity within a node's subgraph, leading to community structure), but also dissortativity (preferential connectivity to a subgraph to which the node does not belong) and core-periphery structure (densely interconnected core and periphery to the core). SBMs can be fitted by statistical inference to an observed graph, or can generate random graphs with predefined structure. \emph{Modularity}, denoted by $Q$, is a specific graph measure derived from stochastic considerations that quantifies density of subgraphs by comparison against a null model:
\begin{equation}
\label{eq_modularity}
Q = \frac{1}{2}\sum_{\substack{i,j\\s_i=s_j}} (a_{i,j}-z_{i,j}) = \frac{1}{2}\sum_{i,j} \left(\frac{1+s_i s_j}{2}\right) (a_{i,j}-z_{i,j}) = \frac{1}{4}\ma s^\top \underbrace{\left(\ma A - \frac{\ma k\ma k^\top}{2M}\right)}_{\ma Q}\ma s,
\end{equation}
where $\ma s$ is a vector encoding the graph partition into two communities, $z_{i,j}=\frac{k_i k_j}{2M}$ is the edge probability between nodes $i$ and $j$ according to the null model, $M = \sum_{i=1}^{N} k_i/2$ is the total edge weight, and $\ma Q$ is the modularity matrix. Choosing this null model allows for comparisons against a reference that preserves the graph degree distribution (i.e., $\sum_{j=1}^{N} a_{i,j}=\sum_{j=1}^{N} z_{i,j}$ for $i=1,2\ldots,N$), with edges placed evenly~\cite{Newman2006}. Hence $Q$ encodes the difference between edge densities in the original graph and in a degree-matched null model. This model is known as the configuration model and is commonly used to define $Q$, but other null models can be considered~\cite{Newman2006,Massen2005,Newman2002assmix}.

The solution to maximizing $Q$ is found by spectral clustering using the eigenvectors of $\ma Q$ with largest eigenvalues, identifying ``modules'' with high assortativity. Similarly, ``anti-modules'' with high dissortativity can be found by minimizing modularity. In fact, $\ma Q$ is a rank-one perturbation of $\ma A$, and consequently, Weyl's inequality informs us that eigenvalues $\lambda_i^{(\ma Q)}$ and $\lambda_i^{(\ma A)}$ of $\ma Q$ and $\ma A$ are interleaved, i.e.,  $\lambda_1^{(\ma A)}\ge \lambda_1^{(\ma Q)}\ge \lambda_2^{(\ma A)}\ge \lambda_2^{(\ma Q)}\ge \ldots \ge \lambda_N^{(\ma A)}\ge \lambda_N^{(\ma Q)}$, where the sequences of eigenvalues are in descending order. This result confirms that the modularity matrix of any simple undirected graph (that is, without self-loops) has both positive and negative eigenvalues~\cite{Bolla2015}. The existence of both positive and negative eigenvalues implies that such a graph can be analyzed in terms of modular and anti-modular spectral components of $\ma Q$. Eigenvectors with zero eigenvalues are modularity-neutral; e.g., the constant vector $\ma 1$ is in the kernel of $\ma Q$ due to $\ma Q \cdot \ma 1 = \ma 0$.

\begin{figure}[!h]
	\centering
	\includegraphics[width=1\textwidth]{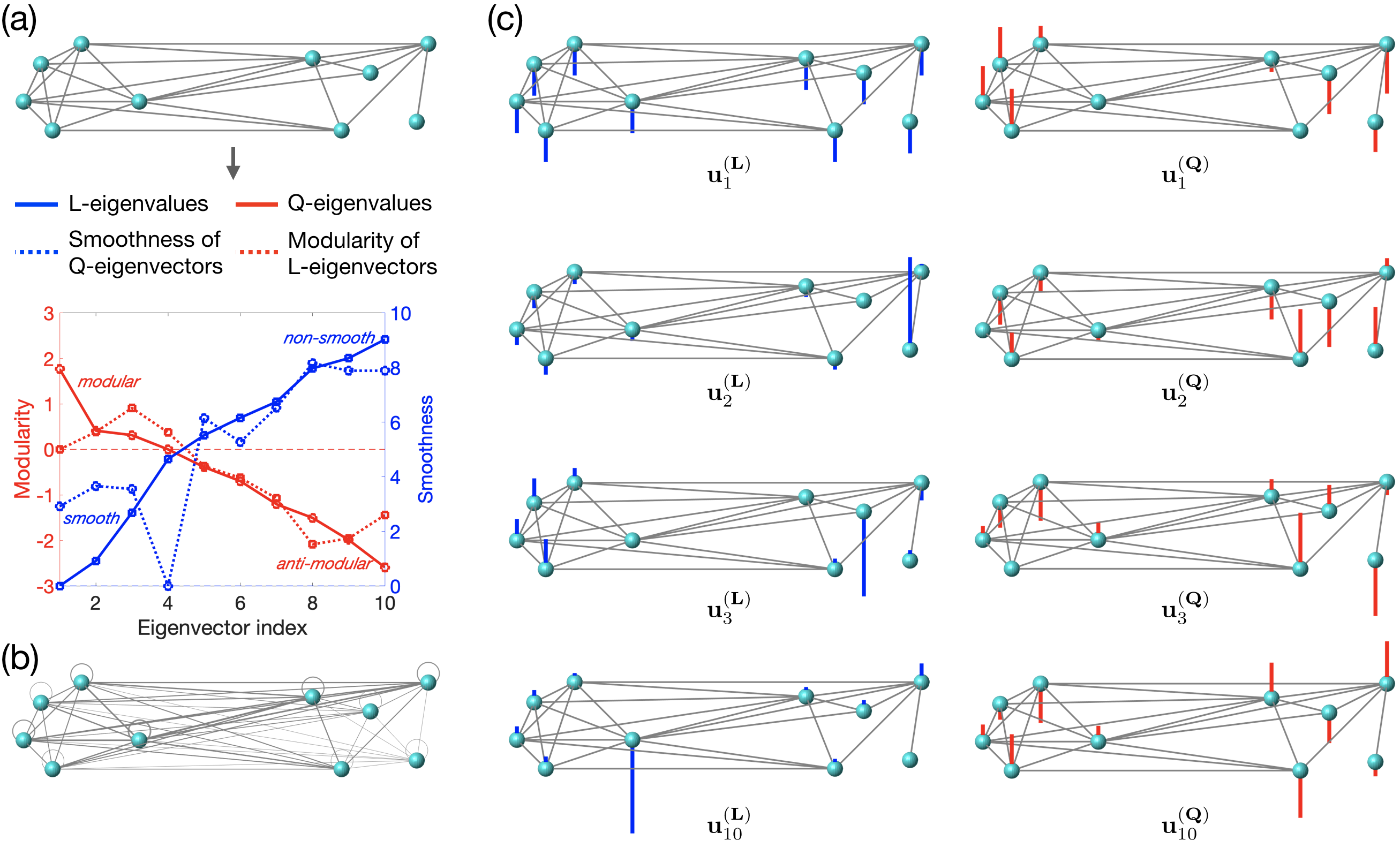}
	\hfil
	\caption{Eigenvalues and eigenvectors of the graph Laplacian ($\ma L$) and modularity matrices ($\ma Q$). (a) Underlying graph structure (top) and corresponding $\ma L$ (blue) and $\ma Q$ (red) eigenspectra (solid lines) and quadratic forms of smoothness and modularity (dotted lines) and (b) corresponding degree-matched null model used to compute $\ma Q$ (Eq. \eqref{eq_modularity}). (c) Selected eigenvectors of $\ma L$ (blue) and $\ma Q$ (red) matrices. Value and sign of eigenvectors' entries are reflected by the height and up-down direction, respectively, of the vertical bars.}
	\label{fig_eigvecs_toy}
\end{figure}

Further illustration of the differences between spectral properties of $\ma L$ and $\ma Q$ is provided in Fig.~\ref{fig_eigvecs_toy} for a toy graph with $N=10$ nodes, $5$ of which form a strong (fully connected) community and the others are weakly connected. Fig.~\ref{fig_eigvecs_toy}a shows the graph and a plot of eigenvalues $\lambda_i^{(\ma L)}$ (blue solid line) and  $\lambda_i^{(\ma Q)}$ (red solid line) in the conventional ascending and descending order, respectively. To better highlight the differences between the corresponding eigenvectors $\ma u_i^{(\ma L)}$ and $\ma u_i^{(\ma Q)}$, respectively, we also plot smoothness ${\ma u_i^{(\ma Q)}}^\top\ma L\ma u_i^{(\ma Q)}$ of modularity eigenvectors (blue dotted line), and modularity\footnote{The quadratic form associated to the modularity matrix will be formally introduced in Sect.~IV.} ${\ma u_i^{(\ma L)}}^\top\ma Q\ma u_i^{(\ma L)}$ of Laplacian eigenvectors (red dotted line). 
Fig.~\ref{fig_eigvecs_toy}b shows the degree-matched null model graph $\frac{\ma k\ma k^\top}{2M}$. Several eigenvectors $\ma u_i^{(\ma Q)}$ and $\ma u_i^{(\ma L)}$ ($i=1,2,3,10$) are shown in Fig.~\ref{fig_eigvecs_toy}c. 
Although $\ma u_i^{(\ma L)}$ are optimized for smoothness, they show high values at specific nodes (except the constant eigenvector $\ma u_1^{(\ma L)}$ with zero eigenvalue). Therefore, the Fiedler vector $\ma u_2^{(\ma L)}$, with lowest graph cut size, does not provide a correct partitioning. The eigenvectors of $\ma Q$ though are optimized for the modularity index and $\ma u_1^{(\ma Q)}$ provides a  conspicuous split between the communities. $\ma Q$ also has a constant eigenvector $\ma u_4^{(\ma Q)}$ with zero smoothness. Curiously, the modularity of $\ma u_3^{(\ma L)}$ is actually the highest among the Laplacian eigenvectors, but still does not provide a convincing partitioning. Eigenvectors of $\ma Q$ with negative eigenvalues, such as $\ma u_{10}^{(\ma Q)}$, are driven by smoothness across modules and signal variability within modules. 

\section{Community-Aware Graph Signal Processing}
\label{sec:GSP_M}

The Laplacian operator $\ma L$ is the common choice of shift operator in GSP~\cite{Shuman2013,Defferrard2016,Huang2018} from which the GFT and all operations are derived. Instead, we propose to use the modularity matrix $\ma Q$ as shift operator. Interestingly, the modularity matrix is a non-local operator since the second term that originates from the null model ``spreads out'' the signal over the whole graph according to the degree distribution---and not only the local neighborhood. Based on this generalized notion of shift operator, we will obtain GSP operations that become aware of the graph community structure, but without the need of explicit community detection. Given a graph signal $\ma x$, its modularity is computed by the quadratic form
\begin{equation}
\label{eq_quadvarQ}
q_Q(\ma x)=\sum_{i,j} a_{i,j}x_ix_j- \frac{\left(\sum_{i,j} a_{i,j}x_i\right)^2}{\sum_{i,j} a_{i,j}}=\ma x^\top \ma Q\ma x.
\end{equation}
Since $\ma Q$ is not positive semi-definite, $q_Q(\ma x)$ can take positive and negative values, depending whether signal variations follow modular or anti-modular organization~\cite{Newman2006}. Thus, the quadratic form $q_Q(\ma x)$ is not a variation norm of the graph signal $\ma x$, which is needed for some GSP operations such as regularization. We overcome this limitation by introducing
\begin{equation}
\label{eq_quadvarQ+}
q_{Q^+}(\ma x) = \ma x^\top \ma Q^+\ma x,
\end{equation}
based on $\ma Q^+=\lambda_{\text{max}}^{(\ma Q)}\cdot \ma I - \ma Q$ where $\lambda_{\text{max}}^{(\ma Q)}$ is the largest eigenvalue of $\ma Q$ and $\ma I$ is the identity matrix. Since $\ma Q^+$ is positive semi-definite, $q_{Q^+}(\ma x)$ is a non-negative function of $\ma x$. A low value of $q_{Q^+}(\ma x)$ reflects that the graph signal $\ma x$ follows modular organization of the graph. On the contrary, a high value of $q_{Q^+}(\ma x)$ is obtained for graph signals reflecting the anti-modular organization. In other words, $q_{Q^+}(\ma x)$ can be interpreted as the modularity-based graph signal variation of $\ma x$. Minimization of this metric is achieved by the eigenvectors of $\ma Q^+$ that define the spectral basis of a GFT exploiting modularity of graph signals. Since $\ma Q^+$ and $\ma Q$ have the same eigenvectors with eigenvalues that are reversed and shifted, the eigenvectors of $\ma Q$ define a proper GFT basis that is built up according to modularity/anti-modularity. Similarly, denoting $\ma Q^-={\ma Q} - \lambda^{(\ma Q)}_{\text{min}} \cdot \ma I$ allows to define a variation norm $q_{Q^-}(\ma x)$ that encodes anti-modular organization in low values, while exploiting the same spectral basis since $\ma Q^-$ also has the same eigenvectors as $\ma Q$.

We now illustrate utility of community-aware GSP tools using data from the OpenFlights Airports Database (\url{https://openflights.org/data.html}) that consists of 3281 airports and 67202 routes (Fig.~\ref{fig_flights_all}a). Graph nodes denote airports that are connected by an undirected binary edge if there exists an airline route between them. Node colors reflect a graph \textit{signal} computed as the sum of both departing and incoming flights at each airport, which was then demeaned and scaled to unit variance. We considered a ground truth community structure based on the continent to which each airport belongs, resulting in a partition of the nodes into the six following communities: Europe, Africa, Asia, Oceania, North America, and South America~\cite{Guimera2005}. The inset of Fig.~\ref{fig_flights_all}a shows the total number of flights leaving from or arriving to the airports in the eastern part of the North American continent. It can be seen that Atlanta airport has more traffic than JFK airport in New-York and that the vast majority of airports have very low traffic.

\subsection{Filtering}
\label{sec:filt}
From the general definition of GSP filtering proposed in Eq.\,\eqref{eq:filtering}, community-aware filtering uses the modularity-based spectral domain with a spectral window $\tilde{\ma h}$:
\begin{equation}
\ma x_\text{out} = \ma U \underbrace{\text{diag}(\tilde{\ma h})}_{=\tilde{\ma H}}\ma U^\top \ma x,
\label{eq_filtering}
\end{equation}
where $\ma U$ contains the eigenvectors of $\ma Q$. While low- and high-pass filtering are natural operations when using the Laplacian GFT, modularity-based GFT allows to define a modular filter (i.e., $\tilde{\ma h}$ has non-zero weights on spectral components with positive eigenvalues) or an anti-modular filter (i.e., $\tilde{\ma h}$ only has non-zero weights on spectral components with negative eigenvalues).

\begin{figure}[!h]
	\centering
	\includegraphics[width=\textwidth]{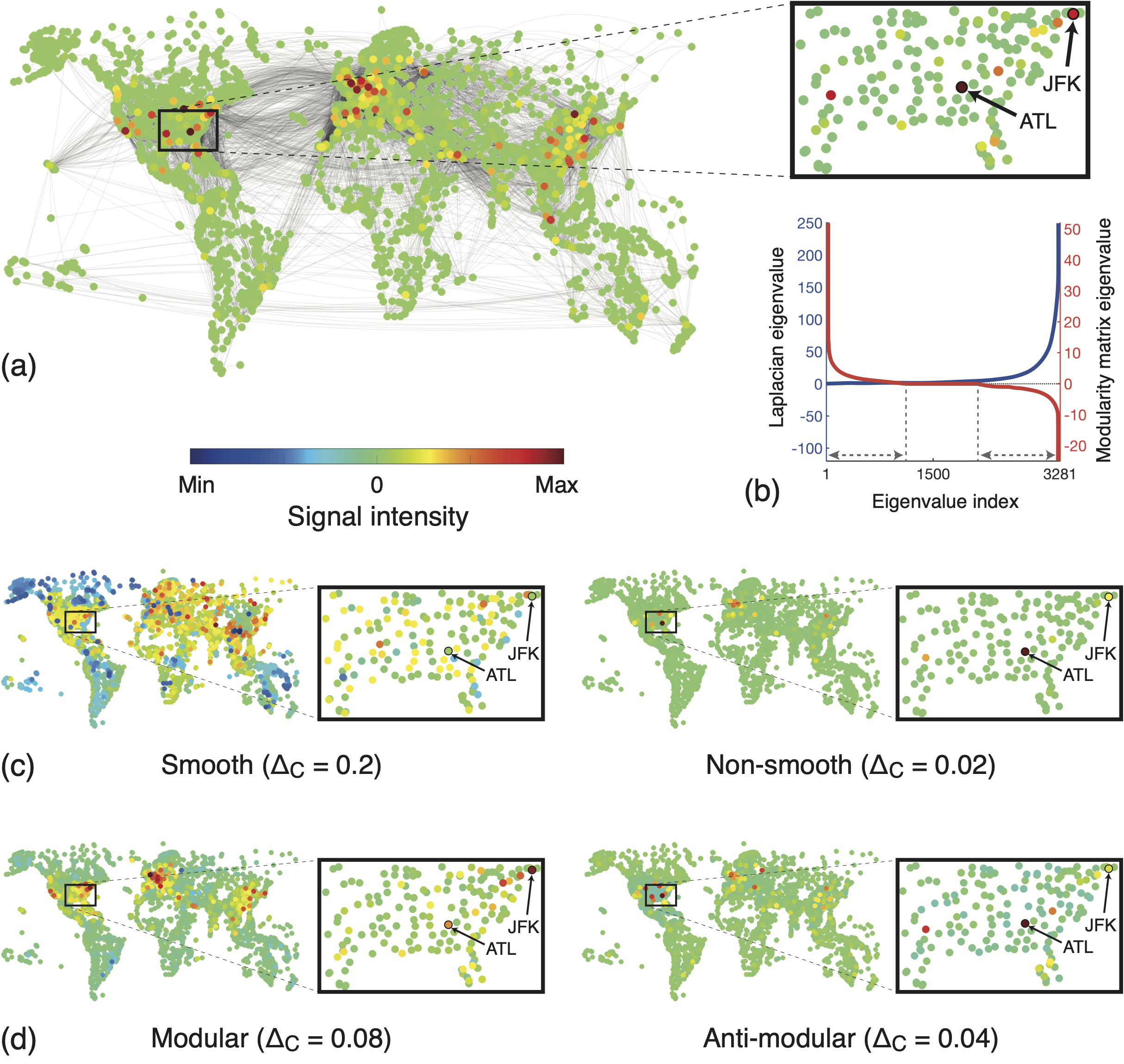}
	\hfil
	\caption{Application of the GSP framework on the OpenFlights. (a) Graph nodes correspond to airports and an edge connects two nodes when at least one flight connects the two corresponding airports. Graph signal (number of flights at each airport) is reflected in nodes' colors. The inset shows a zoom on New-York (JFK) and Atlanta (ATL) airports. (b) Eigenvalues of the graph Laplacian (blue) and modularity matrix (red). Dashed gray lines and arrows represent limits of the filtering passbands, left for modular and smooth, right for anti-modular and non-smooth. (c) Laplacian filtering of the graph signal shown in panel (a) yields smooth and non-smooth signals, and (d) community-aware filtering yields modular and anti-modular signals. The value of within-community variability ($\Delta_C$) is shown for the four filtered signals.}
	\label{fig_flights_all}
\end{figure}

The community-aware filtering was applied on the graph signal of Fig.~\ref{fig_flights_all}a and was compared to a Laplacian-based filtering. The passband, that is, the range of eigenvalue indices with non-zero filter weights, of the modular (anti-modular) filter includes all $1125$ ($1159$) strictly positive (negative) eigenvalues of $\ma Q$, and the smooth (non-smooth) filter was matched so as to capture the same number of spectral Laplacian components (Fig.~\ref{fig_flights_all}b). Within a passband $[N_1,N_2]$, the $i^\text{th}$ entry of $\tilde{\ma h}$ was set to $|\lambda_i|/\sum_{k=N_1}^{N_2} |\lambda_k|$ for modular, anti-modular and non-smooth filterings and to $1-|\lambda_i|/\sum_{k=N_1}^{N_2}|\lambda_k|$ for the smooth filtering, which accounts for strength of modularity or smoothness of the components. Finally, for each filtered signal we computed a measure of within-community variability, denoted $\Delta_C$, and defined as the standard deviation of the filtered signal values within a ground-truth community, averaged over the 6 communities.

As  shown in Fig.~\ref{fig_flights_all}c, the Laplacian-based filtering extracts smooth and non-smooth parts of the graph signal. The smooth signal tends to capture widespread fluctuations over the graph whereas the non-smooth signal contains rather localized  peaks that partially correspond to the extreme values in the original signal (e.g., ATL and JFK airports), suggesting that the underlying community structure is not a predominant feature encoded in Laplacian-filtered signals. In contrast, the modular signal shown in Fig.~\ref{fig_flights_all}d reflects the community structure of the underlying graph by the clusters of high values in North America, Europe and Asia. This effect is further supported by within-community variability ($\Delta_C$) that is lower in the modular signal than in the smooth one. In other words, modular-based filtering can be seen as promoting smoothness \textit{within communities}. On the contrary, the anti-modular signal promotes variability within communities, as this signal shows higher $\Delta_C$ compared to the non-smooth signal. 

In order to further explore the roles of particular nodes in the different filtering operations, we focus on two airports: ATL and JFK. While these two airports are both highly connected, as seen from Fig.~\ref{fig_flights_all}a, they play different roles in the graph community structure. Indeed, ATL has a within-community z-score degree ($Z_{in}$)~\cite{Guimera2005nature} of $8.98$ and an outside-community z-score degree ($Z_{out}$) of $5.97$, while for JFK, we have $Z_{in}=4.22$ and $Z_{out}=15.29$. Therefore, ATL has stronger connections within its community than between communities, and vice versa for JFK. For Laplacian filtering, the signal values of both JFK and ATL are evened out in the smooth signal, and stand out in the non-smooth signal (insets of Fig.~\ref{fig_flights_all}c). However, community-aware filtering picks up differences between these airports by a relatively stronger value of JFK in the modular signal and of ATL in the anti-modular signal (insets of Fig.~\ref{fig_flights_all}d). Since modularity-promoting filtering favors smoothing within the communities, the value of the strongly within-community connected ATL will be more reduced than for JFK. The large value of ATL is captured in the anti-modular signal as it stands out with respect to values of its within-community neighbors. This suggests that modular/anti-modular signal identifies nodes with high values and strong \textit{inter-}modular/\textit{intra-}modular connectivity.
Overall, the results reveal that community-aware filtering can attenuate or enhance values of nodes according to their connectivity within or between communities. 

Finally, the need for the eigendecomposition of $\ma Q$ in Eq.\,\eqref{eq_filtering} can be circumvented by implementing the filtering operation in the vertex domain by a polynomial matrix function $p(\ma Q)$ as suggested in Eq.\,\eqref{eq:filtering}, which is equivalent to applying the spectral window $\tilde{\ma H}=p(\ma\Lambda)$. 
In order to further improve computational efficiency of filtering for large-scale but sparse graphs, one can break down the operation $\ma Q\ma x$ into $\ma Q\ma x=\ma A\ma x-(1/2M)\ma k \ma k^\top \ma x$, where the first term is a sparse matrix-vector multiplication, and the second term can be evaluated by consecutively computing $\ma k^\top \ma x$ and then multiplying the resulting scalar with $\ma k/2M$. Therefore, the dense matrix $\ma Q$ does never need to be stored explicitly. For an undirected graph with $M'$ edges and $N$ nodes, computing $\ma A\ma x$ takes $\mathcal{O}(2M')$ time, and $\ma L\ma x$ takes $\mathcal{O}(2M'+N)$ (in big O notation). The term $(1/2M)\ma k \ma k^\top \ma x$ has complexity of $\mathcal{O}(N)$. Consequently, $\ma Q\ma x$ takes $\mathcal{O}(2M'+N)$, identical to $\ma L\ma x$. For a polynomial filter of order $K$ applied to a large sparse graph, this reverts to $\mathcal{O}(KM')$.

\subsection{Optimal Sampling \& Reconstruction}

Finding the subset of nodes from which a signal can be optimally reconstructed has been extended to the graph domain in the context of bandlimited graph signals $\ma x=\ma B\ma x=\ma U \ma \Sigma \ma U^\top\ma x$~\cite{Tsitsvero.2016}, where $\ma U$ contains the eigenvectors of the shift operator, and $\ma \Sigma$ is a diagonal matrix indicating the passband. The noisy graph signal $\ma y=\ma x + \ma n$, with $\ma n$ additive independent and identically distributed (i.i.d.) noise, is sampled into $\ma x_\text{s}=\ma R\ma y$ where the diagonal matrix $\ma R$ indicates with $0$'s and $1$'s the sampled nodes. Reconstruction denotes the procedure of finding $\ma x_{\text{rec}}$ from $\ma x_\text{s}$, such that the mean squared error $\mathbb{E}[||\ma x_{\text{rec}}-\ma x||_2^2]$ is minimized~\cite{Tsitsvero.2016}. The minimization condition further extends to the choice of \textit{optimal} sampling procedure since sampling at specific nodes can limit the potential performance of the subsequent reconstruction. One of the solutions~\cite{Tsitsvero.2016} to the problem of finding (sub)optimal sampling and reconstruction defines sampling as finding $\ma R^\star $ via:
\begin{equation}
\ma R^\star = \argmax_{\ma R} ||\ma \Sigma \ma U^\top \ma R||_{F}.
\label{eq_subsample}
\end{equation}
Solving Eq.\,\eqref{eq_subsample} amounts to selecting nodes for the optimal sampling subset for which the columns of $\ma \Sigma \ma U^\top$ have the highest $l_2$-norm. Given the graph signal $\ma x_\text{s}$ sampled at nodes defined in $\ma R^\star$, the reconstruction follows:
\begin{equation}
\ma x_{\text{rec}}=\ma V \ma \Psi^{-1} \ma V^\top\ma x_\text{s},
\label{eq_reconstruct}
\end{equation}
\noindent where $\ma V$ and $\ma \Psi$ contain the eigenvectors and eigenvalues of $\ma B\ma R^\star \ma B^\top$.

\begin{figure}[!h]
	\centering
	\includegraphics[width=0.55\textwidth]{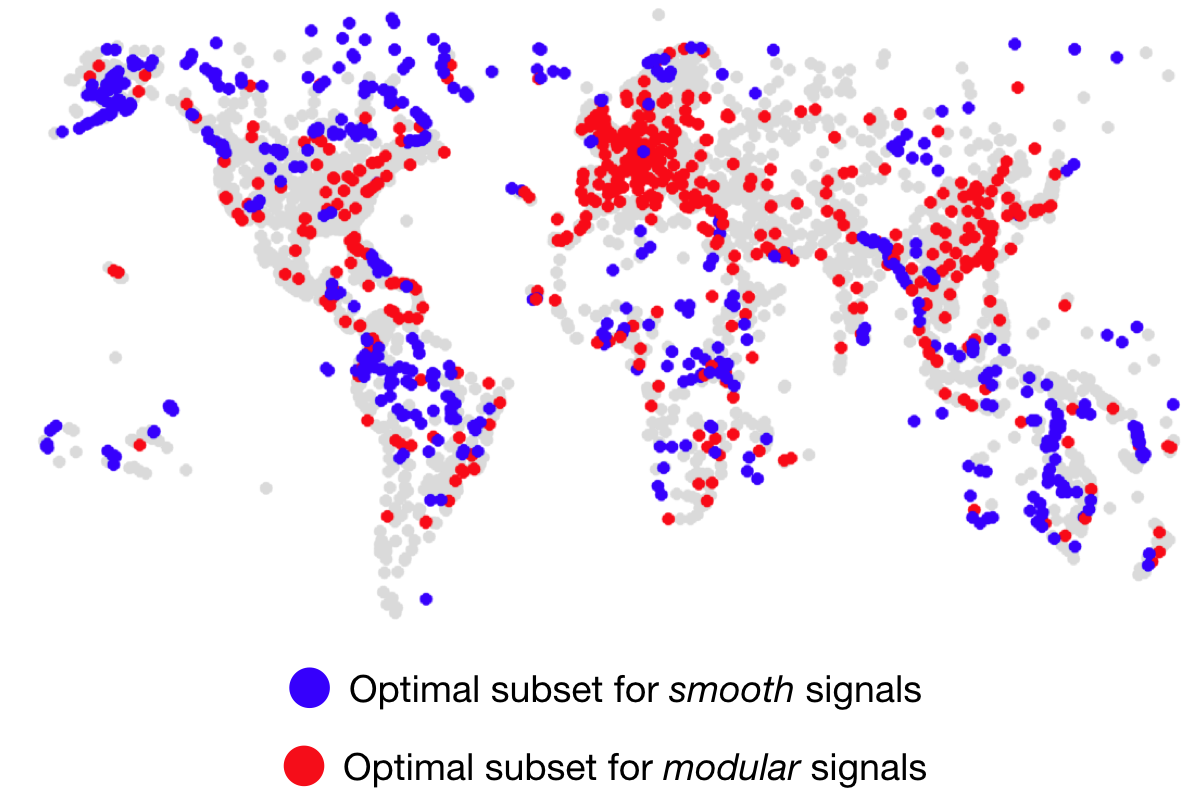}
	\hfil
	\caption{Optimal subset of $500$ nodes for subsampling smooth (blue) or modular (red) signals of bandwidth $200$.}
	\label{fig_optimal_subset}
\end{figure}

We explore how well the graph signal presented in Fig.~\ref{fig_flights_all}a can be reconstructed using the above framework and considering either $\ma L$ or $\ma Q$ as shift operator. We set to $500$ the number of nodes to be sampled and use a spectral band including $200$ components with lowest ($\ma L$), or highest positive ($\ma Q$) eigenvalues. The set of optimal nodes in these two cases is given in Fig.~\ref{fig_optimal_subset}. 

The Laplacian-based sampling subset is composed of more peripheral nodes with low degree ($1.78  \pm  1.01$) whereas the modularity-based sampling contains nodes with high degree ($ 53.12  \pm 43.61$), which are important for inter- and intra-community connectivity. Only two nodes were found to belong to both subsets. An interpretation for this is that while the Laplacian framework focuses on preserving values on nodes where that value is hard to predict due to their low connectivity, the modularity framework maximizes predictability of all nodes by selecting nodes with high degree. This is in accordance with the assumption that traffic at a well connected airport could be a good predictor of the traffic at airports connected to it. Finally, the average reconstruction error is found to be significantly lower ($p<0.01$, paired $t$-test over nodes) for the modular-based framework than for the Laplacian-based framework ($0.0001\pm 0.0002$ vs. $0.0006\pm 0.0037$). This result supports the relevance of the modularity matrix as shift operator rather than the Laplacian in applications where community structure is pertinent.

\subsection{Surrogate Data Generation}

Surrogates play an essential role in non-parametric statistical testing to provide data under the null hypothesis; i.e., randomizing measurements while also preserving some properties. For instance, phase randomization preserves the moduli of the Fourier coefficients while their phases are randomized, leading to surrogate data with the same autocorrelation properties as the original data. This framework was extended to graph signals using the Laplacian, yielding surrogate data that preserve smoothness of the original graph signal~\cite{Pirondini2016}. 
We propose to transpose this method to community-aware representations in order to preserve modular organization of a given graph signal. In particular, a community-aware surrogate signal $\ma x_\text{surr}$ of the graph signal $\ma x$ is given by
\begin{equation}
\ma x_\text{surr}=\ma U \ma C \hat{\ma x},
\label{eq_surrogate}
\end{equation}
where $\hat{\ma x}$ is the modularity-based GFT of $\ma x$ and $\ma C$ a diagonal matrix with random entries $1$ or $-1$, thereby preserving the modularity index of the original signal. The null distribution of any test statistic can then be obtained from multiple realizations of $\ma x_\text{surr}$ and compared against its value for the empirical signal $\ma x$.
This could be refined to a more specific null model by only changing the signs of (anti-)modular components; i.e., entries in $\ma C$ corresponding to positive (negative) eigenvalues of $\ma Q$.

We applied this framework to the signal of Fig.~\ref{fig_flights_all}a by permuting signs of (i) all Laplacian-based, (ii) all modularity-based, (iii) only modular, and (iv) only anti-modular spectral coefficients. For each case, we generated $10000$ surrogate samples that were used to test whether the original signal value is higher than expected under the null hypothesis. The test used an $\alpha$ level of $0.05$ Bonferroni-corrected for multiple comparisons. In (i) and (ii), no nodes were found with values significantly different from their surrogates, but when only randomizing modular (anti-modular) components, 16 (2) airports revealed higher values than expected.

These airports had lower values of $Z_{in}$ ($-0.35 \pm 0.16$) than $Z_{out}$ ($-0.18 \pm 0.03$), indicating these nodes have stronger connectivity with other communities. Considering the results of the filtered signal values of JFK and ATL (Fig.~\ref{fig_flights_all}d), one could expect that high signal values at these nodes can be explained by the underlying community structure. However, surrogate testing showed they cannot be explained by community structure \textit{alone}. Similarly, two `outlier' airports are identified when only randomizing signs of anti-modular spectral coefficients. The results illustrate the complementary roles of modular and anti-modular parts to describe a graph signal. In the context of OpenFlights, this corroborates the assumption of relevant community structure being present in the graph that is only accounted for by modularity-based GSP, and can then be used to assess to what extend graph signals follow this underlying graph structure.

\subsection{Denoising}
Another generic GSP operation is the recovery of the graph signal $\ma x$ from its noisy observation $\ma y=\ma x+\ma n$. The variational formulation puts forward a data-fitting term and a regularization term:
\begin{equation}
\underset{\ma x}{\operatorname{argmin }} \, {||\ma x - \ma y||_2^2 +\mu \cdot \ma x^\top\ma P\ma x },
\label{eq_noisy_rec}
\end{equation}
where $\mu$ is the regularization tuning parameter and the quadratic form of $\ma P$ reflects prior knowledge about $\ma x$. A classical choice is $\ma P=\ma L$, which corresponds to assuming that the graph signal $\ma x$ should be smooth on the graph. Since $\ma L$ is positive semi-definite, the cost function in Eq.\,\eqref{eq_noisy_rec} is convex and has a unique optimal solution. 
The same is true if $\ma P$ is chosen as $\ma Q^+$ and $\ma Q^-$ (cf. Eq.\,\eqref{eq_quadvarQ+}) in order to favor modular or anti-modular organization of $\ma x$, respectively.

Performance of these reconstruction approaches is illustrated using the original signal of Fig.~\ref{fig_flights_all}a. This signal was normalized to unit norm, corrupted with additive Gaussian noise of different variance $\sigma^2$ ranging between $0.01$ and $1$, and the optimal value of $\mu$ was determined using an oracle approach. For small to intermediate noise levels, we found that imposing a modular structure on $\ma x$ (i.e., $\ma P = \ma Q^+$) yielded the best performance (RMS error is $0.0048$ for $\sigma^2 = 0.01$, and $0.0083$ for $\sigma^2 = 0.25$). The error increases by an order of magnitude ($0.0106$ for $\sigma^2 = 0.01$, and $0.0168$ for $\sigma^2 = 0.25$) when using a Laplacian regularizer (i.e., $\ma P = \ma L$), and whereas for anti-modular regularization (i.e., $\ma P = \ma Q^-$) similar values of RMS are reached ($0.0096$ for $\sigma^2 = 0.01$, and $0.0145$ for $\sigma^2 = 0.25$). The advantage of modular regularization decreases for larger noise and the reconstruction errors become comparable when $\sigma^2 = 1$ (all errors above $0.15$).

The assumptions of the different regularizers can be summarized as follows: Laplacian ($\ma L$) favors smoothness of the graph signal by measuring differences between adjacent nodes; modularity ($\ma Q^+$) favors smoothness of the graph signal between nodes weighted by their closeness community-wise (assortative mixing); anti-modularity ($\ma Q^-$) favors smoothness of the graph signal  between nodes weighted inversely by this closeness (dissortative mixing). An explanation of the superior performance of modularity-based regularization in the present example is that similarly high air traffic is more bound to modular organization than to neighborhood relationships; e.g., low traffic of a small island airport connected to several high-traffic mainland hubs of different communities, would lead to a high signal variation through the Laplacian lens, but not through the modularity lens since the island is not close to the large communities. In the end, it is the nature of the graph signal and how it relates to the underlying graph structure that will motivate the use of one or the other regularizer.

\section{Validation for Neuroimaging}
\label{sec:Val}

The results presented above are built from a flight network with known community structure. Likewise, many real-world networks exhibit community structure and we therefore expect the proposed framework to provide a more appropriate way to analyse the corresponding graph signals. We further illustrate the benefits of community-aware GSP in a validation experiment using brain anatomical and functional data from the Human Connectome Project~\cite{VanEssen2013} using a parcellation of the cerebral cortex ($N=360$). The graph \textit{structure} was defined by counting the number of fiber tracts in diffusion-weighted MRI~\cite{Preti.2019}, and the graph \textit{signals} are the activity patterns at different timepoints obtained from functional MRI time series reflecting activity in each brain region~\cite{VanEssen2013}. For each  region, the timecourse was z-scored (centered and unit variance).

The experiment consisted in exploring the link between brain imaging data and $62$ behavioral scores for $181$ healthy volunteers. To that aim, functional time series were filtered using either the anatomical graph Laplacian or modularity matrix following the procedure described in Section~\ref{sec:filt}, and the filtered time series were averaged to yield, for each subject, a metric reflecting (non-)smoothness or (anti-)modular structure of brain function in each brain region~\cite{Medaglia2018}. The link between these measures and the $62$ behavioral scores was then computed using a nested cross-validation scheme and the $R^2$ coefficient of determination was used to quantify the strength of the association between brain function and behavior.

\begin{figure}[!h]
	\centering
	\includegraphics[width=1\textwidth]{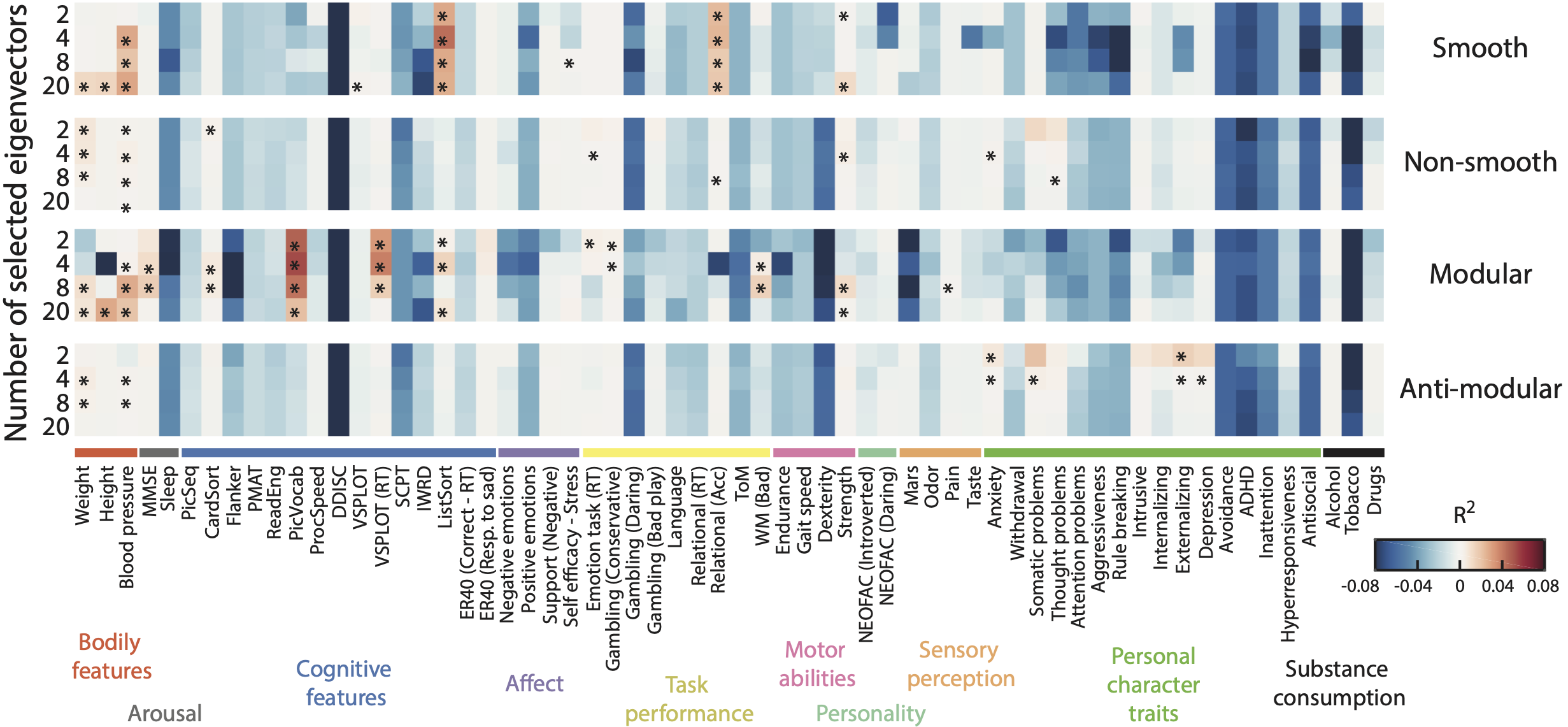}
	\hfil
	\caption{Link between brain function and behavior. Filtering schemes of fMRI time series: smooth and non-smooth using the anatomical graph Laplacian, and modular and anti-modular using the modularity matrix. Results are shown for bandwidths of 2, 4, 8 or 20. The symbol ($^*$) denotes significance on the whole population upon permutation testing ($p<0.01$), with $R^2 > 0$. MMSE: mini mental state examination. PicSeq: picture sequence memory. CardSort: dimensional change card sort. PMAT: Penn progressive matrices. ReadEng: oral reading recognition. PicVocab: picture vocabulary. ProcSpeed: pattern completion processing speed. DDISC: delay discounting. VSPLOT: variable short Penn line orientation test. RT: response time. SCPT: short Penn continuous performance test. Perf.: performance. IWRD: Penn word memory test. ListSort: list sorting. ER40: Penn emotion recognition test. Acc: accuracy. ToM: theory of mind. WM: working memory. Mars: contrast sensitivity. ADHD: attention-deficit hyperactivity disorder.}
	\label{fig_brain}
\end{figure}

Fig.~\ref{fig_brain} shows the values of $R^2$ for the $62$ behavioral measures and different bandwidths. It can first be seen that in most cases using a narrow bandwidth yields stronger $R^2$ which suggests that the information of interest is captured in the very first (non-)smooth or (anti-)modular eigenvectors. Then, we also observe that community-aware filtering reveals links between brain function and behavior that are not captured by Laplacian-based filterings. For example, the anti-modular functional signal shows strong links with various personal character traits while the modular signal mainly captures information about cognitive features.

\section{Conclusion \& Outlook}
Measures of community structure have been extensively used in network science to probe the organization of complex networks. Importantly, the tools that have been developed for processing graph \emph{signals} expressed on these networks are based on the graph Laplacian and thus blind to underlying community structure. We proposed to make GSP community-aware, not by detecting communities, but by defining operations based on the modularity matrix. This provides a natural interpretation of the modularity spectrum in terms of modular and anti-modular contributions, though it requires adaptation when a variation metric is needed. We showed, using several examples, that community-aware GSP acts meaningfully differently compared to classical GSP. Considering the variety of datasets with community structure, the proposed framework will find its use in a wide range of fields and applications.

One extension of Laplacian- and modularity-based GSP is to account for directed graphs. We believe this is beyond the scope of the present paper but the interested reader is referred to \cite{Sardellitti2017} for defining Laplacian-based spectral bases of directed graphs, or modularity matrices for directed graphs~\cite{Malliaros2013} using in- and out-degrees of nodes. Another extension of the community-aware GSP framework could include different null models in the modularity criterion. Specifically, the Bernoulli model preserves the average degree~\cite{Newman2006}, whereas the configuration model can be modified to exclude self-loops~\cite{Massen2005}, or to consider possible correlation between degrees of nodes~\cite{Newman2002assmix}. Finally, communities could also be defined at the level of edges instead of nodes to deal with overlapping communities~\cite{Ahn2010}, or even triangles and higher-order simplicial complexes as in higher-order Laplacian-based topological GSP~\cite{Barbarossa2020}.


%



\section*{Acknowledgment}

This work was supported in part by the Swiss National Science Foundation (200021\_17 5506), the CHIST-ERA IVAN project (20CH21\_174081), the Japan JST ERATO Grant (JPMJER1801), and the Center for Biomedical Imaging (CIBM).

\ifCLASSOPTIONcaptionsoff
  \newpage
\fi



\begin{singlespace}
\bibliographystyle{IEEEtran}
\bibliography{bib30}

\begin{thebibliography}{10}
\providecommand{\url}[1]{#1}
\csname url@samestyle\endcsname
\providecommand{\newblock}{\relax}
\providecommand{\bibinfo}[2]{#2}
\providecommand{\BIBentrySTDinterwordspacing}{\spaceskip=0pt\relax}
\providecommand{\BIBentryALTinterwordstretchfactor}{4}
\providecommand{\BIBentryALTinterwordspacing}{\spaceskip=\fontdimen2\font plus
\BIBentryALTinterwordstretchfactor\fontdimen3\font minus
  \fontdimen4\font\relax}
\providecommand{\BIBforeignlanguage}[2]{{%
\expandafter\ifx\csname l@#1\endcsname\relax
\typeout{** WARNING: IEEEtran.bst: No hyphenation pattern has been}%
\typeout{** loaded for the language `#1'. Using the pattern for}%
\typeout{** the default language instead.}%
\else
\language=\csname l@#1\endcsname
\fi
#2}}
\providecommand{\BIBdecl}{\relax}
\BIBdecl

\bibitem{Shuman2013}
D.~I. Shuman, S.~K. Narang, P.~Frossard, A.~Ortega, and P.~Vandergheynst, ``The
  emerging field of signal processing on graphs: Extending high-dimensional
  data analysis to networks and other irregular domains,'' \emph{{IEEE} Signal
  Processing Magazine}, vol.~30, no.~3, pp. 83--98, May 2013.

\bibitem{Sandryhaila2013}
A.~Sandryhaila and J.~M.~F. Moura, ``Discrete signal processing on graphs,''
  \emph{{IEEE} Transactions on Signal Processing}, vol.~61, no.~7, pp.
  1644--1656, Apr. 2013.

\bibitem{Hammond2011}
D.~K. Hammond, P.~Vandergheynst, and R.~Gribonval, ``Wavelets on graphs via
  spectral graph theory,'' \emph{Applied and Computational Harmonic Analysis},
  vol.~30, no.~2, pp. 129--150, Mar. 2011.

\bibitem{Tsitsvero.2016}
M.~Tsitsvero, S.~Barbarossa, and P.~Di~Lorenzo, ``Signals on graphs:
  Uncertainty principle and sampling,'' \emph{IEEE Transactions on Signal
  Processing}, vol.~64, no.~18, pp. 4845--4860, 2016.

\bibitem{Fortunato2016}
\BIBentryALTinterwordspacing
S.~Fortunato and D.~Hric, ``Community detection in networks: A user guide,''
  \emph{Physics Reports}, vol. 659, pp. 1--44, Nov. 2016. [Online]. Available:
  \url{https://doi.org/10.1016/j.physrep.2016.09.002}
\BIBentrySTDinterwordspacing

\bibitem{Newman2004}
\BIBentryALTinterwordspacing
M.~E.~J. Newman, ``Fast algorithm for detecting community structure in
  networks,'' \emph{Physical Review E}, vol.~69, no.~6, June 2004. [Online].
  Available: \url{https://doi.org/10.1103/physreve.69.066133}
\BIBentrySTDinterwordspacing

\bibitem{Jeub2015}
\BIBentryALTinterwordspacing
L.~G.~S. Jeub, P.~Balachandran, M.~A. Porter, P.~J. Mucha, and M.~W. Mahoney,
  ``Think locally, act locally: Detection of small, medium-sized, and large
  communities in large networks,'' \emph{Physical Review E}, vol.~91, no.~1,
  Jan. 2015. [Online]. Available:
  \url{https://doi.org/10.1103/physreve.91.012821}
\BIBentrySTDinterwordspacing

\bibitem{Guimera2005}
\BIBentryALTinterwordspacing
R.~Guimera, S.~Mossa, A.~Turtschi, and L.~A.~N. Amaral, ``The worldwide air
  transportation network: Anomalous centrality, community structure, and
  cities' global roles,'' \emph{Proceedings of the National Academy of
  Sciences}, vol. 102, no.~22, pp. 7794--7799, May 2005. [Online]. Available:
  \url{https://doi.org/10.1073/pnas.0407994102}
\BIBentrySTDinterwordspacing

\bibitem{Guimera2005nature}
\BIBentryALTinterwordspacing
R.~Guimer{\`{a}} and L.~A.~N. Amaral, ``Functional cartography of complex
  metabolic networks,'' \emph{Nature}, vol. 433, no. 7028, pp. 895--900, Feb.
  2005. [Online]. Available: \url{https://doi.org/10.1038/nature03288}
\BIBentrySTDinterwordspacing

\bibitem{Meunier2010}
D.~Meunier, R.~Lambiotte, and E.~T. Bullmore, ``Modular and hierarchically
  modular organization of brain networks,'' \emph{Frontiers in Neuroscience},
  vol.~4, 2010.

\bibitem{Newman2006}
\BIBentryALTinterwordspacing
M.~E.~J. Newman, ``Finding community structure in networks using the
  eigenvectors of matrices,'' \emph{Physical Review E}, vol.~74, no.~3, Sep.
  2006. [Online]. Available: \url{https://doi.org/10.1103/physreve.74.036104}
\BIBentrySTDinterwordspacing

\bibitem{vonLuxburg2007}
U.~von Luxburg, ``A tutorial on spectral clustering,'' \emph{Statistics and
  Computing}, vol.~17, no.~4, pp. 395--416, Aug. 2007.

\bibitem{Ortega2018}
A.~Ortega, P.~Frossard, J.~Kova{\v c}evi{\'c}, J.~M.~F. Moura, and
  P.~Vandergheynst, ``Graph signal processing: Overview, challenges, and
  applications,'' \emph{Proceedings of the IEEE}, vol. 106, no.~5, pp.
  808--828, May 2018.

\bibitem{Marques2017}
A.~G. Marques, S.~Segarra, G.~Leus, and A.~Ribeiro, ``Stationary graph
  processes and spectral estimation,'' \emph{{IEEE} Transactions on Signal
  Processing}, vol.~65, no.~22, pp. 5911--5926, Nov. 2017.

\bibitem{Defferrard2016}
M.~Defferrard, X.~Bresson, and P.~Vandergheynst, ``Convolutional neural
  networks on graphs with fast localized spectral filtering,'' in
  \emph{Advances in Neural Information Processing Systems 29}, D.~D. Lee,
  M.~Sugiyama, U.~V. Luxburg, I.~Guyon, and R.~Garnett, Eds.\hskip 1em plus
  0.5em minus 0.4em\relax Curran Associates, Inc., 2016, pp. 3844--3852.

\bibitem{Huang2018}
\BIBentryALTinterwordspacing
W.~Huang, T.~A.~W. Bolton, J.~D. Medaglia, D.~S. Bassett, A.~Ribeiro, and
  D.~V.~D. Ville, ``A graph signal processing perspective on functional brain
  imaging,'' \emph{Proceedings of the {IEEE}}, vol. 106, no.~5, pp. 868--885,
  May 2018. [Online]. Available:
  \url{https://doi.org/10.1109/jproc.2018.2798928}
\BIBentrySTDinterwordspacing

\bibitem{leonardi1302}
N.~Leonardi and D.~Van De~Ville, ``Tight wavelet frames on multislice graphs,''
  \emph{IEEE Transactions on Signal Processing}, vol.~61, no.~13, pp.
  3357--3367, 2013.

\bibitem{Tremblay.2014}
N.~Tremblay and P.~Borgnat, ``Graph wavelets for multiscale community mining,''
  \emph{IEEE Transactions on Signal Processing}, vol.~62, no.~20, pp.
  5227--5239, 2014.

\bibitem{Lee2019}
C.~Lee and D.~J. Wilkinson, ``A review of stochastic block models and
  extensions for graph clustering,'' \emph{Applied Network Science}, vol.~4,
  no.~1, Dec. 2019.

\bibitem{Massen2005}
C.~P. Massen and J.~P.~K. Doye, ``Identifying communities within energy
  landscapes,'' \emph{Physical Review E}, vol.~71, no.~4, Apr. 2005.

\bibitem{Newman2002assmix}
M.~E.~J. Newman, ``Assortative mixing in networks,'' \emph{Physical Review
  Letters}, vol.~89, no.~20, Oct. 2002.

\bibitem{Bolla2015}
M.~Bolla, B.~Bullins, S.~Chaturapruek, S.~Chen, and K.~Friedl, ``Spectral
  properties of modularity matrices,'' \emph{Linear Algebra and its
  Applications}, vol. 473, pp. 359--376, May 2015.

\bibitem{Pirondini2016}
E.~Pirondini, A.~Vybornova, M.~Coscia, and D.~V.~D. Ville, ``A spectral method
  for generating surrogate graph signals,'' \emph{{IEEE} Signal Processing
  Letters}, vol.~23, no.~9, pp. 1275--1278, Sep. 2016.

\bibitem{VanEssen2013}
D.~C.~V. Essen, S.~M. Smith, D.~M. Barch, T.~E. Behrens, E.~Yacoub, and
  K.~Ugurbil, ``The {WU}-minn human connectome project: An overview,''
  \emph{{NeuroImage}}, vol.~80, pp. 62--79, Oct. 2013.

\bibitem{Preti.2019}
M.~G. Preti and D.~V.~D. Ville, ``Decoupling of brain function from structure
  reveals regional behavioral specialization in humans,'' \emph{Nature
  Communications}, vol.~10, no.~1, Oct. 2019.

\bibitem{Medaglia2018}
J.~D. Medaglia, W.~Huang, E.~A. Karuza, A.~Kelkar, S.~L. Thompson-Schill,
  A.~Ribeiro, and D.~S. Bassett, ``Functional alignment with anatomical
  networks is associated with cognitive flexibility,'' \emph{Nature Human
  Behaviour}, vol.~2, no.~2, pp. 156--164, 2018.

\bibitem{Sardellitti2017}
S.~Sardellitti, S.~Barbarossa, and P.~D. Lorenzo, ``On the graph fourier
  transform for directed graphs,'' \emph{{IEEE} Journal of Selected Topics in
  Signal Processing}, vol.~11, no.~6, pp. 796--811, Sep. 2017.

\bibitem{Malliaros2013}
F.~D. Malliaros and M.~Vazirgiannis, ``Clustering and community detection in
  directed networks: A survey,'' \emph{Physics Reports}, vol. 533, no.~4, pp.
  95--142, Dec. 2013.

\bibitem{Ahn2010}
Y.-Y. Ahn, J.~P. Bagrow, and S.~Lehmann, ``Link communities reveal multiscale
  complexity in networks,'' \emph{Nature}, vol. 466, no. 7307, pp. 761--764,
  Aug. 2010.

\bibitem{Barbarossa2020}
S.~Barbarossa and S.~Sardellitti, ``Topological signal processing over
  simplicial complexes,'' \emph{{IEEE} Transactions on Signal Processing}, pp.
  1--1, Mar. 2020.

\end{thebibliography}
\end{singlespace}
%


%






\end{document}